\documentclass[aps,twocolumn]{revtex4}

\usepackage{color}
\usepackage{graphicx}
\usepackage{amsmath}
\usepackage{amssymb}
\usepackage[]{hyperref}

\usepackage{ulem}

\renewcommand{\emph}[1]{{\it #1}}

\begin{document}

\title{Efficient frequency conversion based on resonant four-wave mixing}

\author{Chin-Yao Cheng$^{1}$}
\author{Zi-Yu Liu$^{1}$}
\author{Pi-Sheng Hu$^{1}$}
\author{Tsai-Ni Wang$^{1}$}
\author{Chung-Yu Chien$^{1}$}
\author{Jia-Kang Lin$^{1}$}
\author{Jz-Yuan Juo$^{1}$}
\author{Jiun-Shiuan Shiu$^{1}$}
\author{Ite A. Yu$^{2,5}$}
\author{Ying-Cheng Chen$^{3,5}$}
\author{Yong-Fan Chen$^{1,4,5}$}

\email{yfchen@mail.ncku.edu.tw}

\affiliation{
$^1$Department of Physics, National Cheng Kung University, Tainan 70101, Taiwan \\
$^2$Department of Physics, National Tsing Hua University, Hsinchu 30013, Taiwan \\
$^3$Institute of Atomic and Molecular Sciences, Academia Sinica, Taipei 10617, Taiwan \\
$^4$Center for Quantum Frontiers of Research \& Technology, Tainan 70101, Taiwan \\
$^5$Center for Quantum Technology, Hsinchu 30013, Taiwan
}



\begin{abstract}

Efficient frequency conversion of photons has important applications in optical quantum technology because the frequency range suitable for photon manipulation and communication usually varies widely. Recently, an efficient frequency conversion system using a double-$\Lambda$ four-wave mixing (FWM) process based on electromagnetically induced transparency (EIT) has attracted considerable attention because of its potential to achieve a nearly 100\% conversion efficiency (CE). To obtain such a high CE, the spontaneous emission loss in this resonant-type FWM system must be  suppressed considerably. A simple solution is to arrange the applied laser fields in a backward configuration. However, the phase mismatch due to this configuration can cause a significant decrease in CE. Here, we demonstrate that the phase mismatch can be effectively compensated by introducing the phase shift obtained by two-photon detuning. Under optimal conditions, we observe a wavelength conversion from 780 to 795 nm with a maximum CE of 91.2$\% $ $\pm$ 0.6$\% $ by using this backward FWM system at an optical depth of 130 in cold $^{87}$Rb atoms. The current work represents an important step toward achieving low-loss, high-fidelity EIT-based quantum frequency conversion.


\end{abstract}




\maketitle


\newcommand{\FigOne}{
    \begin{figure}[t]
    \centering
    \includegraphics[width=9.0 cm]{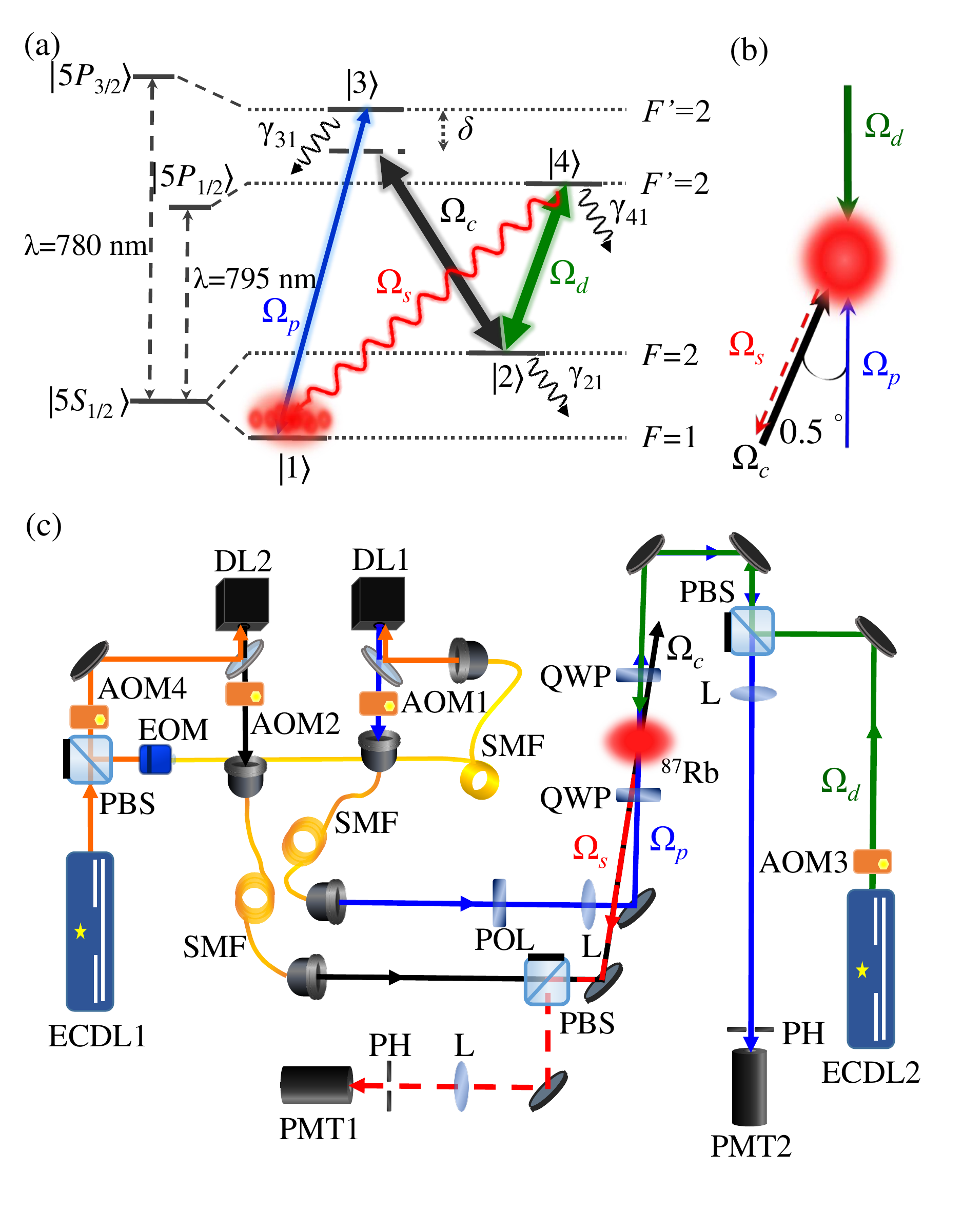}
    \caption{
Energy level diagram and experimental settings. (a) Energy levels of the $^{87}$Rb $D_1$ and $D_2$ lines used in the backward FWM experiment. Here, two-photon detuning, $\delta$, has the same magnitude and opposite direction as does coupling detuning. Coupling detuning is defined as $\omega_c-\omega_{23}$, where $\omega_c$ and $\omega_{23}$ are the frequencies of the coupling field and the $|2\rangle\leftrightarrow|3\rangle$ transition, respectively. (b) Angle set between the light fields, plotted on the upper right. (c) Schematic of the apparatus setup. ECDL, external cavity diode laser; DL, diode laser; EOM, electro-optic modulator; AOM, acousto-optic modulator; SMF, single-mode fiber; POL, polarizer; PBS, polarization beam splitter; QWP, quarter-wave plate; L, lens; PH, pinhole; PMT, photo-multiplier tube.
}
    \end{figure}
}


\newcommand{\FigTwo}{
    \begin{figure}[t]
    \includegraphics[width=9 cm]{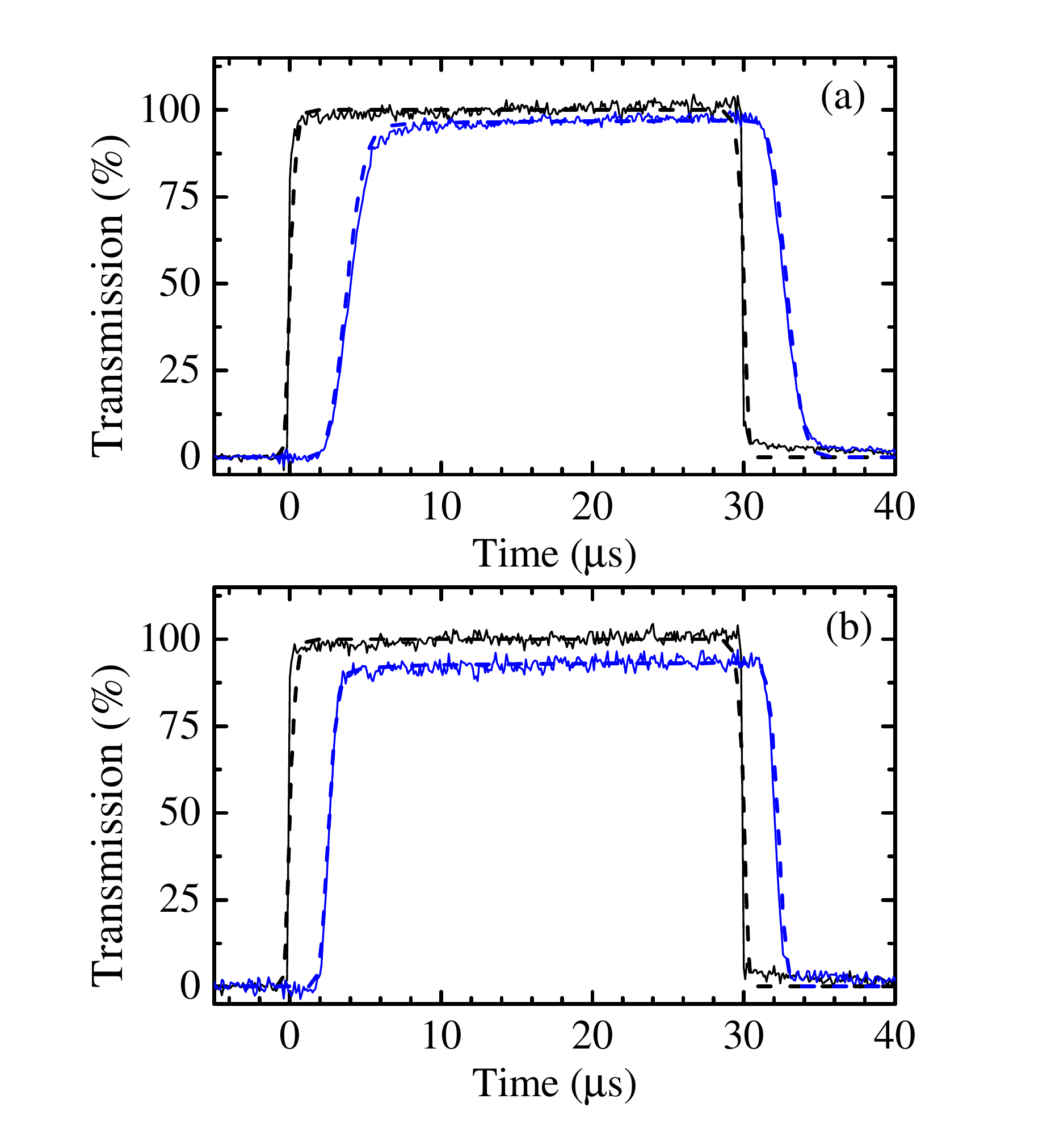}
    \caption{
Observation of the slow light effect based on EIT. The black (higher) and blue (lower) lines show the incident probe pulse and the probe pulse propagating through the medium, respectively. The solid lines represent the experimental data, whereas the dashed lines represent the theoretical curves with the following parameters: (a) $\alpha=45$, $\Omega_c=0.60\Gamma$, and $\gamma_{21}=2\times 10^{-4}\Gamma$; (b) $\alpha=130$, $\Omega_c=1.20\Gamma$, and $\gamma_{21}=7\times 10^{-4}\Gamma$.
}
    \end{figure}
}


\newcommand{\FigThree}{
    \begin{figure}[t]
    \includegraphics[width=9 cm]{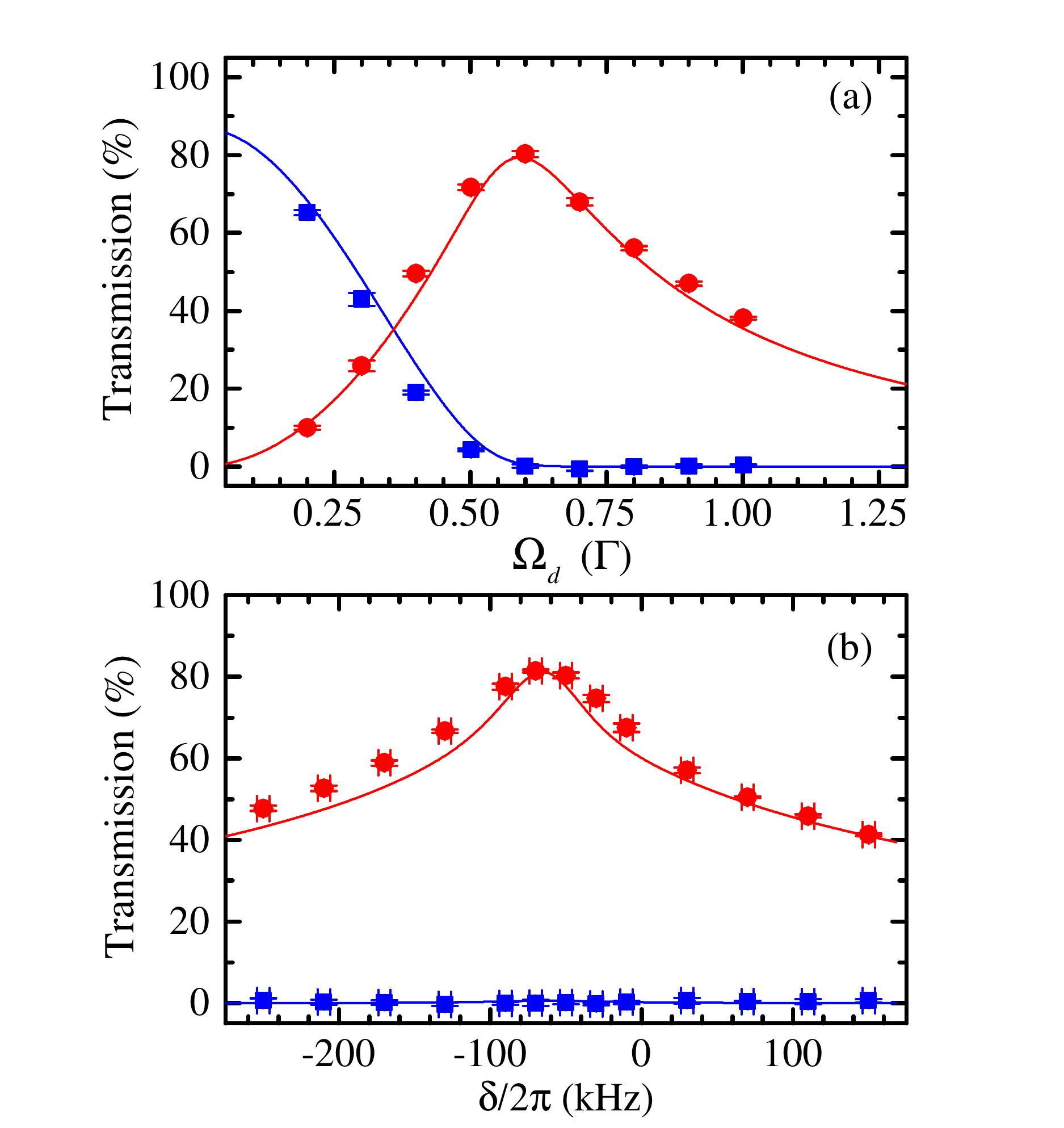}
    \caption{
Observation of the backward FWM process in the MOT. With the change of the driving Rabi frequency (a) and two-photon detuning (b), the steady-state transmittance of the probe  and signal pulses propagating in the FWM medium are measured simultaneously. The blue squares and red circles represent the probe and signal transmissions, respectively. The solid lines show the theoretical curves calculated by Eqs. (1)--(5) with the parameters in (a) $\alpha=45$, $\Omega_c=0.60\Gamma$, $\gamma_{21}=2\times 10^{-4} \Gamma$, $\Delta kL=0.447\pi$, and $\delta/2\pi$ = --54 kHz; (b) $\alpha=45$, $\Omega_c=\Omega_d=0.60\Gamma$, $\gamma_{21}=2\times 10^{-4} \Gamma$, and $\Delta kL=0.447\pi$. The experimental data show that the maximum CE of the backward FWM reaches 81.4$\%$ $\pm$ 0.3$\%$.
}
    \end{figure}
}


\newcommand{\FigFour}{
    \begin{figure}[t]
    \includegraphics[width=9 cm]{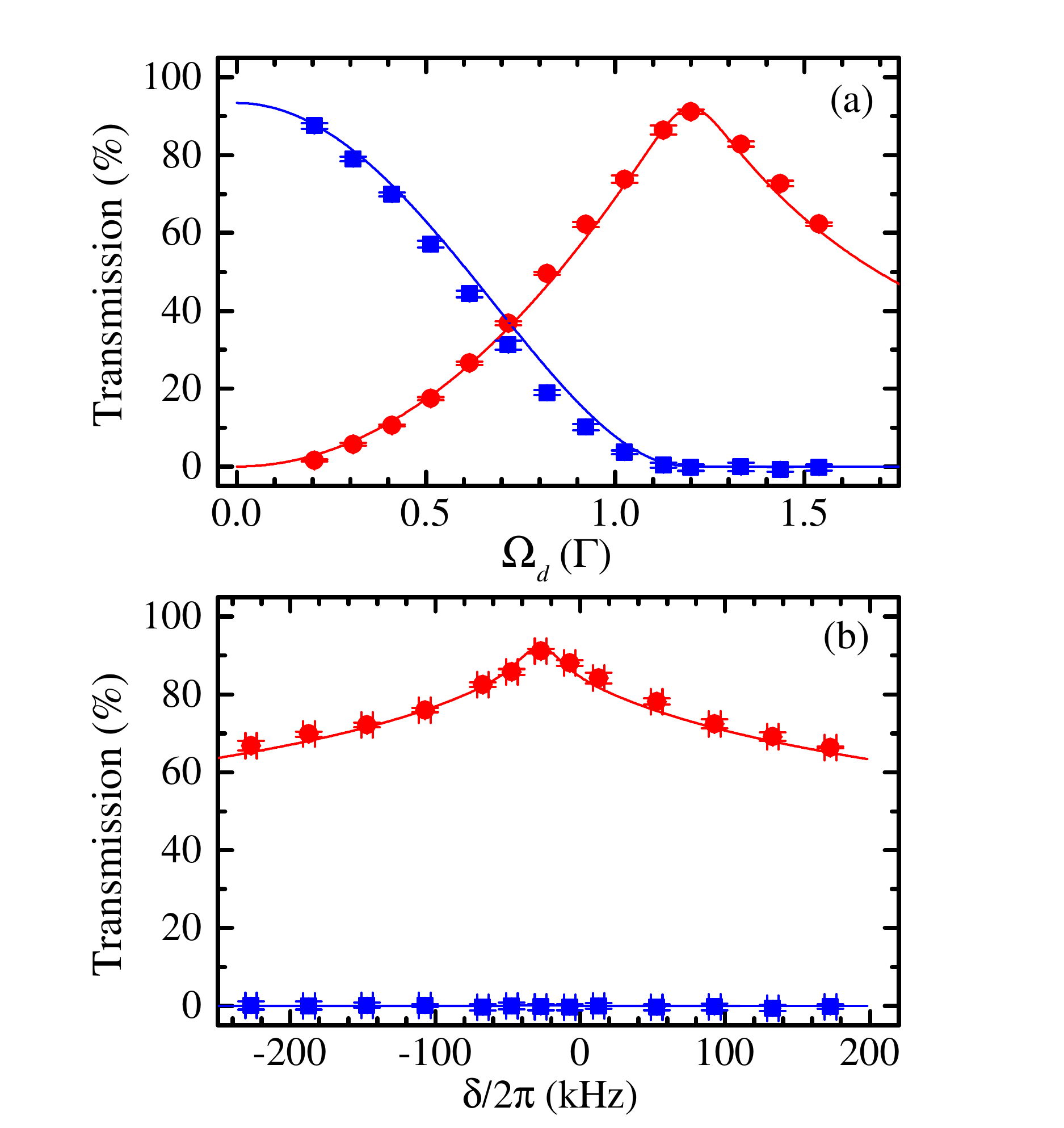}
    \caption{ 
Observation of the backward FWM process in the dark SPOT. With the change of the driving Rabi frequency (a) and two-photon detuning (b), the steady-state transmittance of the probe and signal pulses propagating in the FWM medium are measured simultaneously. The blue squares and red circles represent the probe and signal transmissions, respectively. The solid lines show the theoretical curves calculated by Eqs. (1)--(5) with the parameters in (a) $\alpha=130$, $\Omega_c=1.20\Gamma$, $\gamma_{21}=7\times 10^{-4} \Gamma$, $\Delta kL=0.134\pi$, and $\delta/2\pi$ = --27 kHz; (b) $\alpha=130$, $\Omega_c=\Omega_d=1.20\Gamma$, $\gamma_{21}=7\times 10^{-4} \Gamma$, and $\Delta kL=0.134\pi$. The experimental data show that the maximum CE of the backward FWM reaches 91.2$\%$ $\pm$ 0.6$\%$.
}
    \end{figure}
}


\newcommand{\FigFive}{
    \begin{figure}[t]
    \includegraphics[width=9 cm]{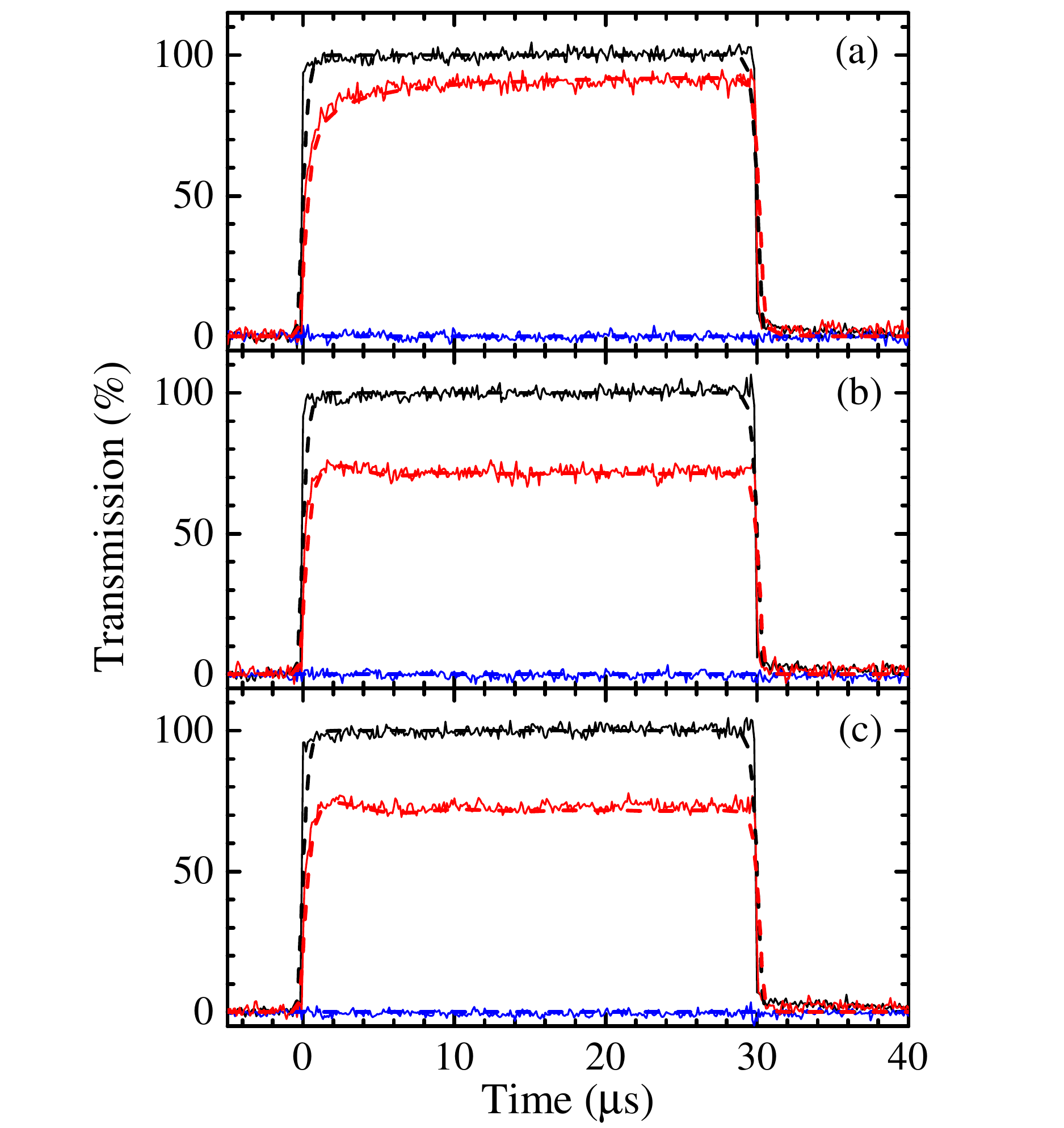}
    \caption{ 
Observation of the backward FWM in the pulsed regime. The black (higher), red (middle), and blue (lower) lines show the incident probe, the generated signal, and the transmitted probe pulses, respectively. The solid lines represent the experimental data, whereas the dashed lines are the theoretical curves with the following parameters: $\alpha=130$, $\Omega_c=\Omega_d=1.20 \Gamma$, $\gamma_{21}=7\times 10^{-4} \Gamma$, $\Delta kL=0.134\pi$, and (a) $\delta/2\pi$ = --27 kHZ; (b) $\delta/2\pi$ = 93 kHz; (c) $\delta/2\pi$ = --147 kHz. The results in (a) shows that the phase shift due to two-photon detuning in the backward FWM system effectively compensates for the phase mismatch and achieves a maximum CE of approximately 91\%. The data in (b) and (c) show that the phase mismatch is not well compensated, resulting in a significant decrease in CE.
}
    \end{figure}
}


\section{Introduction}

Quantum frequency conversion of photons is important in the development of long-distance quantum communication and effective optical quantum computing~\cite{QFC by Kumar, DLCZ, QC by Zeilinger}. In nonlinear optics, the wave mixing effect away from resonance can not only prevent the quantum noise caused by vacuum fluctuations but also convert the photon frequency to a larger range and bandwidth~\cite{Boyd NLO}. Therefore, in the past few decades, researchers have often used the wave mixing effect far from resonance to achieve efficient frequency conversion. Most of the experiments to achieve high-efficiency frequency conversion are performed in solid materials along with sum-frequency generation~\cite{SFG1, SFG2, SFG4, SFG6} or Bragg-scattering four-wave mixing (FWM)~\cite {BS1, BS2, BS3, BS4}. In nonlinear optical systems, the highest internal conversion efficiency (CE) has reached thus far has been of \textgreater90\% by using sum frequency mixing in a nonlinear crystal and Bragg-scattering FWM in an optical fiber~\cite{SFG1, BS4}. In nonlinear crystals, a strong pump light is usually required to achieve a high-efficiency QFC. However, under strong pump light conditions, additional noise photons are often generated due to spontaneous Raman or parametric conversion effects, which can cause difficulties in the practical application of QFC~\cite{Parametric Noise}. Although the required pump power can be reduced by using waveguides, cavities or fibers, this results in the coupling loss of incident photons, thereby reducing the overall CE of QFC.

Another promising approach for implementing low-loss frequency converters is the use of a resonant FWM system based on  electromagnetically induced transparency (EIT)~\cite{Harris FWM2, Kang BFWM, Yu FWM, Liu BFWM, Juo FWM}. Because EIT can greatly enhance nonlinear interactions between photons and considerably suppress vacuum field noise under ideal conditions, many EIT-based quantum applications have been proposed and demonstrated at the single-photon level; these applications include quantum memory~\cite{Hsiao QM, Laurat QM, Zhu QM}, photonic transistors~\cite{PTransister1, PTransister2, PTransister3}, optical phase gates~\cite{XPM2, XPM3, XPM4}, and frequency beam splitter~\cite{Yu FBS}. An efficient FWM based on double-$\Lambda$ EIT system has attracted much attention recently because it can achieve nearly 100\% CE and can protect the quantum state of frequency-converted photons~effectively~\cite{Chen QFC}. To obtain such a high CE, effective suppression of the spontaneous emission loss in this resonant-type FWM system is essential. A simple solution involves backward configuration of the applied laser fields~\cite{Kang BFWM, Liu BFWM}. However, the phase mismatch due to the backward configuration can cause significant decreases in CE. Here, we demonstrate that the phase mismatch can be effectively compensated by introducing a phase shift obtained through two-photon detuning. Under optimal conditions, we observe a wavelength conversion from 780 to 795 nm with a maximum CE of 91.2$\%$ $\pm$ 0.6$\%$ by using this backward FWM system in cold rubidium ($^{87}$Rb) atoms. To our knowledge, this is the highest CE that has been achieved in an EIT-based FWM system thus far.

\FigOne 


\section{Experiment Details}

Here, we investigate a resonant-type frequency conversion based on backward FWM in a Doppler-cooled $^{87}$Rb atomic system. The arrangement of the light fields and relative energy levels are shown in Fig.~1(a). The probe field ($\Omega_p$ indicates its Rabi frequency) with an intensity of approximately 1 $\mu$W/cm$^2$ drives the transition of the state $|1\rangle\leftrightarrow|3\rangle$ and forms a $\Lambda$-type EIT system alongside the coupling field ($\Omega_c$), which drives the transition of the state $|2\rangle\leftrightarrow|3\rangle$ with an intensity of approximately 10 mW/cm$^2$. Under the EIT condition, the driving field ($\Omega_d$) drives the transition of the state $|2\rangle \leftrightarrow |4\rangle$ and induces the FWM process to generate the signal field ($\Omega_s$) during the transition of the state $|1\rangle \leftrightarrow |4\rangle$. The wavelength consequently converts from 780 nm ($\Omega_p$) to 795 nm ($\Omega_s$). All the light fields are right-circularly polarized ($\sigma+$), and the spatial configuration of these light fields is shown in Fig.~1(b). The angle between the probe field and the coupling field is set at 0.5$^\circ$ to suppress unwanted light from leaking into the photon detector. The driving field is counter-propagating to the probe field as a backward FWM condition. Because of the phase match condition, the generated signal field is counter-propagating to the coupling field. Notably, this backward configuration causes a non-negligible phase mismatch effect and significantly reduces CE. However, as long as the two-photon resonance of the built-in EIT mechanism (formed by $\Omega_p$ and $\Omega_c$) in the backward FWM system is slightly destroyed to cause a  phase shift in the probe field, this phase mismatch effect can be effectively compensated.

Figure~1(c) illustrates the experimental setup. The probe and coupling fields are produced from the diode laser 1 (DL1) and DL2, respectively. The coupling field is directly injection locked by the external cavity diode laser 1 (ECDL1). One beam from the ECDL1 is sent through a 6.8-GHz electro-optic modulator (EOM). The probe field is injection locked by an intermediate laser seeded with the high-frequency sideband of the EOM output. The above arrangement can eliminate the influence of the EOM output carrier on DL1. Both the probe and coupling fields are sent to their respective single-mode fibers (SMFs) to obtain spatial mode matching. The driving field is generated from ECDL2, another ECDL. All these light fields are turned on and off by their respective acousto-optic modulators (AOMs). We use AOM1 to control the probe pulse width. The coupling and driving fields are turned on and off via AOM2 and AOM3, respectively. AOM4 is used to change the frequency of the coupling field and thereby adjust the two-photon detuning in the FWM experiment.

The 1/e$^{2}$ diameter of the probe beam waist is approximately 300 $\mu$m, whereas the diameters of the coupling beam and driving beam are 3 and 4 mm, respectively. The probe and signal pulses are detected by a photo-multiplier tube module 1 (PMT1) and PMT2, respectively. In the backward FWM experiment, the approximate overall detection efficiencies for the probe and signal fields are 65\% and 75\%, respectively.

Regarding the timing sequence arrangement, we denote the time of firing the probe pulse as $t = 0$ ms. The magnetic field of the magneto-optical trap (MOT) is first switched off at $t = -1.6$ ms. Next, the repumping laser of the MOT is turned off at $t = -200~\mu$s and the coupling and driving fields are turned on immediately. Before sending the probe pulse, the MOT's trapping laser is turned off at $t=-100~\mu$s. This timing arrangement of the optical fields ensures that the entire atomic population is prepared at the ground state $|1\rangle$; moreover, the repetition rate for each measurement is maintained at 100 Hz throughout the experiment. In addition, to ensure that the probe pulse is long enough to reach steady-state conditions, the spectrum width of the probe pulse is required to be smaller than the width of the EIT transparent window; therefore, the duration of the probe pulse is set to 30~$\mu$s in this experiment.


\FigTwo

\section{Theoretical Model}

To theoretically analyze the propagation behavior of the probe and signal pulse in the backward FWM medium, we use Maxwell--Schr\"{o}dinger equations (MSEs):
\begin{eqnarray}
\frac{\partial\Omega_p}{\partial z}+\frac{1}{c}\frac{\partial \Omega_p}{\partial t}=i\frac{\alpha_p \gamma_{31}}{2L}\rho_{31},\\
-\frac{\partial\Omega_{s}}{\partial z}+\frac{1}{c}\frac{\partial\Omega_{s}}{\partial t}-i\Delta k\Omega_{s}=i\frac{\alpha_s \gamma_{41}}{2L}\rho_{41},
\end{eqnarray}
where $\rho_{ij}$ is the slowly varying amplitude of the coherence between $|i\rangle$ and $|j\rangle$; $\gamma_{31}$ and $\gamma_{41}$ represent the total coherence decay rates from the excited states $|3\rangle$ and $|4\rangle$, respectively; $\Delta k=k_p-k_c+k_d-k_s$ is the phase-mismatch term (in which the plus and minus signs indicate the processes of light field absorption and emission by the medium, respectively); $\alpha_{p(s)}=n \sigma_{13(14)} L$ represents the optical depth (OD) of the probe (signal) transition (where $n$ and $L$ are the density and length of the medium, respectively). In this experiment, $\sigma_{13}\approx \sigma_{14}$, because the Clebsch--Gordan coefficients of the probe and signal transitions are symmetric when considering the degenerate Zeeman sublevels; moreover, their wavelengths (780 and 795 nm, respectively) and the spontaneous decay rates ($2\pi \times 6.06$ and $2\pi \times 5.75$ MHz, respectively) are also close to each other. Thus, for simplicity, we assume $\alpha_p=\alpha_s=\alpha$ and $\gamma_{31}=\gamma_{41}$. Notably, $\gamma_{31}$ and $\gamma_{41}$ in this experiment are mainly attributed to the spontaneous decay rate, and both are represented by $\Gamma \approx 2\pi \times 6$ MHz.

To solve MSEs, we use the optical Bloch equations (OBEs) to obtain the equations of motion of $\rho_{31}$ and $\rho_{41}$. Considering the case where both the probe and signal fields are weak enough to be considered the perturbation fields (i.e., $\rho_{11}\approx 1$), the density matrix elements with slowly varying amplitudes of the OBEs are given by
\begin{eqnarray}
\frac{d}{dt}\rho_{21}=\frac{i}{2}\Omega^{\ast}_c\rho_{31}+\frac{i}{2}\Omega^{\ast}_d\rho_{41}+(i\delta-\frac{\gamma_{21}}{2})\rho_{21},\\
\frac{d}{dt}\rho_{31}=\frac{i}{2}\Omega_p+\frac{i}{2}\Omega_c\rho_{21}+(i\Delta_p-\frac{\gamma_{31}}{2})\rho_{31},\\
\frac{d}{dt}\rho_{41}=\frac{i}{2}\Omega_s+\frac{i}{2}\Omega_d\rho_{21}+(i\Delta-\frac{\gamma_{41}}{2})\rho_{41},
\end{eqnarray}
where $\gamma_{21}$ is the dephasing rate between ground states $|1\rangle$ and $|2\rangle$, $\Delta_p=\omega_p-\omega_{31}$ represents one-photon detuning, $\delta=(\omega_p-\omega_c)-\omega_{21}$ is two-photon detuning, and $\Delta=(\omega_p-\omega_c+\omega_d)-\omega_{41}$ is three-photon detuning. We determine the value of each parameter in Eqs (1)--(5) from additional experiments as follows: $\Omega_c$ is measured on the basis of the separation of the two absorption peaks in the EIT spectrum. OD is determined using the delay time of the slow light pulse. $\Omega_d$ is obtained by simulating the generated signal pulse in the backward FWM system.

According to the theoretical predictions, when the Rabi frequencies of the coupling and driving fields are the same, the backward FWM system can achieve the highest CE~\cite{Liu BFWM}. Therefore, for simplicity, we assume the conditions of $\Omega_c=\Omega_d=\Omega$, $\Delta_p = \Delta=0 $, and $\gamma_{21}=0$. By the boundary conditions of $\Omega_p(z=0)=\Omega_{p0}$ and $\Omega_s(z=L)=0$, we solve Eqs. (1)--(5) and obtain the steady-state solutions for the probe and signal fields:

\begin{eqnarray}
\Omega_p(z=L)=\frac{2q}{2q \cos(\beta/2)+\kappa \sin(\beta/2)}\Omega_{p0}e^{-i\Delta kL/2},\\
\Omega_s(z=0)=\frac{\alpha}{\kappa+2q\cot(\beta/2)}\Omega_{p0},
\end{eqnarray}
where
\begin{eqnarray}
\kappa=\left(\alpha-\frac{2\Delta kL\delta\Gamma}{|\Omega|^2}\right)-i2\xi,\\
\beta=\sqrt{\frac{(\Delta kL+i\alpha)(\Delta kL\delta\Gamma+i|\Omega|^2\xi)}{i|\Omega|^2+\delta\Gamma}},\\
q=\sqrt{\left(\xi-\frac{i\Delta kL\delta\Gamma}{|\Omega|^2}+i\alpha\right)\left(\xi-\frac{i\Delta kL\delta\Gamma}{|\Omega|^2}\right)},\\
\xi=\Delta kL+\frac{\delta\alpha\Gamma}{|\Omega|^2}.
\end{eqnarray}
Notably, the parameter $\xi$ can be used to estimate the amount of two-photon detuning needed to compensate for the phase-mismatch effect. In addition, from Eq. (7), the steady-state CE of the backward FWM process can be determined:
\begin{eqnarray}
\texttt{CE}=\left|\frac{\Omega_s(z=0)}{\Omega_{p0}}\right|^2 = \left|\frac{\alpha}{\kappa+2q \cot(\beta/2)}\right|^2.
\end{eqnarray}
If both $\delta$ and $\Delta kL$ are equal to 0, then Eq.~(12) here can be simplified to Eq.~(32) in the reference~\cite{Chen QFC}, which describes a backward FWM system under phase matching and two-photon resonance.

\FigThree

\FigFour


\section{Results and Discussion}

Before performing the backward FWM experiment, we measure the slow light effect based on EIT (Fig. 2). The OD in Fig. 2(a) measured in a typical MOT is approximately 45. To further increase the OD, we measure the slow light  pulse in a dark spontaneous-force optical trap (SPOT). Details on the similar experimental setup of the dark SPOT can be found in our previous study~\cite{Lo XPM}. The OD in the dark SPOT is measured to be approximately 130 [Fig. 2(b)]. Moreover, the measured ground-state dephasing rate $\gamma_{21}$ is dissimilar, which is 2$\times 10^{-4}\Gamma$ and 7$\times 10^{-4}\Gamma$ in Figs. 2(a) and 2(b), respectively. This is mainly because the coupling field interacts with the excited state of not only  $ | 5P_ {3/2}, F = 2 \rangle $ but also the excited state of $ | 5P_ {3/2}, F = 3 \rangle $, which leads to the AC Stark energy shift of the ground state $ | 2 \rangle $. The energy shift of the ground state destroys the two-photon resonance condition of EIT and thus causes the probe field to become absorbed by the EIT medium; this is similar to a nonresonant photon switching effect~\cite{Harris PS}. Although this ground state energy shift can be effectively compensated by introducing detuning of the coupling field, it may still damage the EIT slightly and cause the ground state dephasing rate to increase effectively. To compensate for the energy shift of the ground state caused by the photon switching effect, the detuning of the coupling field in Figs.~2(a) and 2(b) is set to --27 and --106 kHz, respectively. According to theoretical estimates, the ground state dephasing rate due to the photon switching effect increases by approximately 1$\times 10^{-4}\Gamma$ and 5$\times 10^{-4}\Gamma$ in Figs.~2(a) and 2(b), respectively. This is consistent with our experimental observations.

\FigFive

Next, we turn on the driving field to conduct the backward FWM experiment. Figure 3 depicts the experimental observation of the backward FWM effect in MOT. Throughout the backward FWM experiment, five measurement samples are used to calculate the statistical error bars. Each measurement sample averages 1024 times. We measure the steady-state transmittance of the probe and signal pulses propagating in the FWM medium versus the driving Rabi frequency and two-photon detuning, as shown in Fig.~3(a) and 3(b), respectively. According to the spatial configuration of the probe, coupling, and driving fields mentioned in Fig.~1(b), the phase-mismatch parameter $\Delta k$ is calculated to be approximately 0.893$\pi$ cm$^{-1}$.  Because the diameter of the cold atom cloud in the MOT is approximately 5 mm, the value of $\Delta kL$ determined in the experiment is 0.447$\pi$. Figure 3(b) shows that the maximum CE of the incident probe light converted into FWM signal light reaches 81.4$\%$ $\pm$ 0.3$\%$, where the two-photon detuning is set to --70 kHz. 

To further increase the CE, we perform the backward FWM experiment in the dark SPOT. By adjusting the intensity ratio of the six trapping laser beams in the dark SPOT, we generate a cold atom cloud with an OD of up to 130 but with a small diameter ($\approx$1.5 mm). Figure 4 shows the experimental observation in the dark SPOT. Because the value of $\Delta kL$ here is smaller than that at MOT, we use smaller two-photon detuning to compensate the phase-mismatch effect. The experimental data in Fig.~4(b) show that at a two-photon detuning setting of --27 kHz, the maximum CE can reach 91.2$\%$ $\pm$ 0.6$\%$. Moreover, the bandwidth of this resonant backward FWM is approximatively 0.8 MHz, which is mainly affected by the EIT effect and the intensity balance condition~\cite{Liu BFWM}, so the bandwidth can be increased by increasing the intensity of the coupling and driving light.

Figures 5(a), 5(b), and 5(c) present parts of the raw data in Fig.~4(b), where two-photon detuning is --27, 93, and --147 kHz, respectively. On the basis of the experimental conditions, we estimate that the phase shift of the probe field caused by the detuned EIT effect is approximately --0.129$\pi$, 0.405$\pi$, and --0.704$\pi$ in 5(a), 5(b), and 5(c), respectively.  Because the phase mismatch $\Delta k L$ in this backward FWM experiment is approximately 0.134$\pi$, it can be well compensated by the two-photon detuning condition in Fig.~5(a). By contrast, poor phase matching conditions cause a significant reduction in CE, as shown in Figs.~5(b) and 5(c).


\section{Conclusion}
Here, we demonstrated the use of EIT-based double-$\Lambda$ FWM mechanism for efficient frequency conversion. The spatial backward arrangement of the applied light fields can suppress the spontaneous emission loss of the resonant-type FWM system considerably. In this study, we use the small two-photon detuning of about tens of kHz to compensate for the phase-mismatch effect in the backward FWM so as to achieve quasi-phase matching conditions, which can significantly increase the CE of the EIT-based frequency converter. We consequently observe a wavelength conversion from 780 to 795 nm with a 91.2$\%$ CE at an OD of 130 in cold $^{87}$Rb atoms. This low-loss resonant FWM based on EIT provides a promising method for implementing a high-fidelity quantum frequency converter~\cite{Chen QFC} and therefore has potential applications in optical quantum information technology.


\section*{ACKNOWLEDGMENTS}

We thank Pei-Chen Kuan, Wen-Te Liao, and Ray-Kuang Lee for their useful contributions to discussions. This work was supported by the Ministry of Science and Technology of Taiwan [grant numbers: 107-2112-M-006-008-MY3 and 108-2639-M-007-001-ASP]. We also acknowledge the support from Center for Quantum Technology in Taiwan.




\begin{thebibliography}{1}

\bibitem{QFC by Kumar}
    P. Kumar,
    Opt. Lett. \textbf{15}, 1476 (1990).

\bibitem{DLCZ}  
    L.-M. Duan, M. D. Lukin, J. I. Cirac, and P. Zoller,
    Nature (London) \textbf{414}, 413 (2001). 

\bibitem{QC by Zeilinger}  
    N. K. Langford, S. Ramelow, R. Prevedel, W. J. Munro, G. J. Milburn, and A. Zeilinger,
    Nature (London) \textbf{478}, 360 (2011).

\bibitem{Boyd NLO}  
    R. W. Boyd, \textit{Nonlinear Optics}, 3rd ed. (San Diego, Academic, 2008).

\bibitem{SFG1}
    M. A. Albota and F. N. C. Wong,
    Opt. Lett. \textbf{29}, 1449 (2004)
    
\bibitem{SFG2}  
    S. Tanzilli, W. Tittel, M. Halder, O. Alibart, P. Baldi, N. Gisin, and H.  Zbinden,
    Nature (London) \textbf{437}, 116 (2005).

\bibitem{SFG4}
    R. Ikuta, Y. Kusaka, T. Kitano, H. Kato, T. Yamamoto, M. Koashi, and N. Imoto,
    Nat. Commun. \textbf{2}, 537 (2011).

\bibitem{SFG6}
    N. Maring, D. Lago-Rivera, A. Lenhard, G. Heinze, and H. de Riedmatten,
    Optica \textbf{5}, 507 (2018).

\bibitem{BS1}
    H. J. McGuinness, M. G.  Raymer, C. J. McKinstrie, and S. Radic, 
    Phys. Rev. Lett. \textbf{105}, 093604 (2010).

\bibitem{BS2}
    A. S. Clark, S. Shahnia, M. J. Collins, C. Xiong, and B. J. Eggleton,
    Opt. Lett. \textbf{38}, 947 (2013).

\bibitem{BS3}
    Q. Li, M. Davanço, and K. Srinivasan,
    Nat. Photon. \textbf{10}, 406 (2016).

\bibitem{BS4}
    S. Clemmen, A. Farsi, S. Ramelow, A. Gaeta, 
    Phys. Rev. Lett. \textbf{117}, 223601 (2016).
    
\bibitem{Parametric Noise}
    J. S. Pelc, C. Langrock, Q. Zhang, and M. M. Fejer,
    Opt. Lett. \textbf{35}, 2804 (2010).

\bibitem{Harris FWM2}
    A. J. Merriam, S. J. Sharpe, M. Shverdin, D. Manuszak, G.Y. Yin, and S. E. Harris,
    Phys. Rev. Lett. \textbf{84}, 5308 (2000).

\bibitem{Kang BFWM}
    H. Kang, G. Hernandez, and Y. Zhu,
    Phys. Rev. A \textbf{70}, 061804(R) (2004).

\bibitem{Yu FWM}
    C.-Y. Lee, B.-H. Wu, G. Wang, Y.-F. Chen, Y.-C. Chen, I. A. Yu,
    Opt. Express \textbf{24}, 1008 (2016).

\bibitem{Liu BFWM}  
    Z.-Y. Liu, J.-T. Xiao, J.-K. Lin, J.-J. Wu, J.-Y. Juo, C.-Y. Cheng, and Y.-F. Chen,
    Sci. Rep. \textbf{7}, 15796 (2017).

\bibitem{Juo FWM}
    J.-Y. Juo, J.-K. Lin, C.-Y. Cheng, Z.-Y. Liu, I. A. Yu, and Y.-F. Chen,
    Phys. Rev. A \textbf{97}, 053815 (2018).

\bibitem{Hsiao QM}
    Y.-F. Hsiao, P.-J. Tsai, H.-S. Chen, S.-X. Lin, C.-C. Hung, C.-H. Lee, Y.-H. Chen, Y.-F. Chen, I. A. Yu, and Y.-C. Chen,
    Phys. Rev. Lett. \textbf{120}, 183602 (2018).

\bibitem{Laurat QM}
    P. Vernaz-Gris, K. Huang, M. Cao, A. S. Sheremet, and J. Laurat,
    Nat. Commun. \textbf{9}, 363 (2018).

\bibitem{Zhu QM}
    Y. Wang, J. Li, S. Zhang, K. Su, Y. Zhou, K. Liao, S. Du, H. Yan, and S.-L. Zhu,
    Nat. Photon. \textbf{13}, 346 (2019).

\bibitem{PTransister1}
    W. Chen, K. M. Beck, R. B\"{u}cker, M. Gullans, M. D. Lukin, H. Tanji-Suzuki, and V. Vuleti\'{c},  
    Science \textbf{341}, 768 (2013).

\bibitem{PTransister2}
    H. Gorniaczyk, C. Tresp, J. Schmidt, H. Fedder, and S. Hofferberth,
    Phys. Rev. Lett. \textbf{113}, 053601 (2014).

\bibitem{PTransister3}
    D. Tiarks, S. Baur, K. Schneider, S. D\"{u}rr, and G. Rempe,
    Phys. Rev. Lett. \textbf{113}, 053602 (2014).

\bibitem{XPM2}
    K. M. Beck, M. Hosseini, Y. Duan, and V. Vuleti\'{c},
    Proc. Natl. Acad. Sci. U.S.A. \textbf{113}, 9740 (2016).

\bibitem{XPM3}
    D. Tiarks, S. Schmidt, G. Rempe, and S. D\"{u}rr,
    Science Adv. \textbf{2}, 4 (2016).

\bibitem{XPM4}
    Z.-Y. Liu, Y.-H. Chen, Y.-C. Chen, H.-Y. Lo, P.-J. Tsai, I. A. Yu, Y.-C. Chen, and Y.-F. Chen,
    Phys. Rev. Lett. \textbf{117}, 203601 (2016).

\bibitem{Yu FBS}
    K.-F. Chang, T.-P. Wang, C.-Y. Chen, Y.-H. Chen, Y.-S. Wang, Y.-F. Chen, Y.-C. Chen, and I. A. Yu,
    arXiv:1907.03393

\bibitem{Chen QFC}
    C.-Y. Cheng, J.-J. Lee, Z.-Y. Liu, J.-S. Shiu, and Y.-F. Chen,
    arXiv:2009.11021
    
\bibitem{Lo XPM}
    H.-Y. Lo, Y.-C. Chen, P.-C. Su, H.-C. Chen, J.-X. Chen, Y.-C. Chen, I. A. Yu, and Y.-F. Chen,
    Phys. Rev. A \textbf{83}, 041804(R) (2011).

\bibitem{Harris PS}
    S. E. Harris and Y. Yamamoto,
    Phys. Rev. Lett. \textbf{81}, 3611 (1998).

\end{thebibliography}
\end{document}